\documentclass[fleqn,twoside]{article}

\def\be{\begin{equation}}
\def\ee{\end{equation}}
\def\ai{\'{\i}}

\def\tp1{\tilde p_1}
\def\tp2{\tilde p_2}
\def\tq1{\tilde q^1}
\def\tq2{\tilde q^2}
\def\om{\Omega}

\newcommand{\Title}[1]{\noindent {\uppercase{\Large #1}} \\}
\newcommand{\Authors}[4]{\noindent
  {\large\bf #1\dag\ #2\ddag}\medskip\begin{description}
  \item[\dag]{\it #3} \item[\ddag]{\it #4}\end{description}}
\newcommand{\Abstract}[1]{\vskip 2mm \begin{center}
  \parbox{16.4cm}{\small\noindent #1} \end{center}\medskip}
\newcommand{\foom}[1]{\protect\footnotemark[#1]}
\newcommand{\email}[2]{\footnotetext[#1]{e-mail: #2}
  \addtocounter{footnote}{1}}

\makeatletter
\renewcommand{\@oddhead}{\raisebox{0pt}[\headheight][0pt]{%
   \vbox{\hbox to\textwidth{{\protect\small\it %
   The Problem of Time and Gauge Invariance -- I}%
   \hfil \rm \thepage \strut}\hrule}}}
\renewcommand{\@evenhead}{\raisebox{0pt}[\headheight][0pt]{%
   \vbox{\hbox to\textwidth{\thepage \hfil {\protect\small\it %
   T. P. Shestakova and C. Simeone} \strut}\hrule}}}
\renewcommand{\section}{\@startsection{section}{1}{0pt}%
  {-3.5ex plus -1ex minus -.2ex}{2.3ex plus .2ex}%
  {\large\bf\protect\raggedright}}

\renewcommand{\subsection}{\@startsection{subsection}{2}{0pt}%
  {-3ex plus -1ex minus -.2ex}{1.4ex plus .2ex}%
  {\normalsize\bf\protect\raggedright}}

\makeatother

\begin{document}
\twocolumn[
\thispagestyle{empty}
\bigskip

\Title{THE PROBLEM OF TIME AND GAUGE INVARIANCE \\[5pt]
       IN THE QUANTIZATION OF COSMOLOGICAL MODELS. \\[5pt]
       I. CANONICAL QUANTIZATION METHODS}

\Authors{T. P. Shestakova\foom 1}
        {and C. Simeone\foom 2}
        {Department of Theoretical and Computational Physics,\\
         Rostov State University, Sorge Str. 5, Rostov-on-Don, 344090, Russia}
        {Departamento de F\'{\i }sica,
         Facultad de Ciencias Exactas y Naturales Universidad de Buenos Aires,\\
         Ciudad Universitaria Pabell\'{o}n I - 1428, Buenos Aires, Argentina\\
         and Instituto de Astronom\ai a y F\ai sica del Espacio
         C.C. 67, Sucursal 28 - 1428 Buenos Aires, Argentina}

\Abstract
{The paper is the first of two parts of a work reviewing some
approaches to the problem of time in quantum cosmology, which were
put forward last decade, and which demonstrated their relation to
the problems of reparametrization and gauge invariance of quantum
gravity. In the present part we remind basic features of quantum
geometrodynamics and minisuperspace cosmological models, and
discuss fundamental problems of the Wheeler -- DeWitt theory.
Various attempts to find a solution to the problem of time are
considered in the framework of the canonical approach. Possible
solutions to the problem are investigated making use of
minisuperspace models, that is, systems with a finite number of
degrees of freedom. At the same time, in the last section of the
paper we expand our consideration beyond the minisuperspace
approximation and briefly review promising ideas by Brown and
Kucha\v r, who propose that dust interacting only gravitationally
can be used for time measuring, and the unitary approach by
Barvinsky and collaborators. The latter approach admits both the
canonical and path integral formulations and anticipates the
consideration of recent developments in the path integral approach
in the second part of our work.}]

\email 1 {shestakova@phys.rsu.ru}
\email 2 {csimeone@df.uba.ar}
\date{}

\section{Introduction}

It is generally accepted now that initial stages of cosmological
evolution must be described by quantum cosmology. The need for
a quantum theory of the early universe is a logical consequence of
the fact that classical General Relativity is not applied in the
vicinity of cosmological singularity. As was pointed out
by Grishchuk and Zeldovich \cite{grish}, a full cosmological
theory must include a notion about the origin of spacetime itself
which is essentially a quantum gravitational phenomenon. In the
framework of such a full theory one should consider both
gravitational field and matter quantized.

The standard approach to quantum cosmology includes three basic
steps: a classical theory for the dynamics, a quantization
prescription in terms of a wave function or a propagator, and
interpretation. The second and third steps are highly non trivial
because General Relativity includes the general covariance as a
central feature. Accordingly, the Hamiltonian formulation for the
gravitational field is that of a constrained system. Any attempt to
save gauge invariance in quantum theory of gravity creates a number
of problems.

The problem of time is the most known difficulty of the Wheeler --
DeWitt quantum geometrodynamics which is a theoretical basis for
modern quantum cosmology. This problem is indissolubly connected
with other ones among which are problems of Hilbert space
(positive-definite inner product), reparametrization noninvariance
and operator ordering. The problem of time has been discussed in a
plenty of papers (see, for example, \cite{vil86, cast88, cast89,
cast90, cast98, fumo89, unruh89a, unruh89b, unruh95, ma95, isham92,
buish99}). The paper by Vilenkin \cite{vil86} was one of the first
works where the problem of time was considered in the context of
quantum cosmology. In the paper by Unruh \cite{unruh89a} it was
shown that a solution of this problem may require some modification
of the theory of gravity (including the Hamiltonian constraint).
Isham \cite{isham92} gave a very informative and profound review of
the problem, a classification of existing approaches to the problem
of time and many references can be found therein. Philosophical
aspects of the problem were discussed in \cite{unruh95, buish99}.

It is not the purpose of the present paper to give exhaustive
consideration to all approaches to the problem of time which are
widespread in modern literature. We do not also intend to repeat
earlier papers on this subject. Our aim is to review some ideas
put forward last decade and show that the problem of time is
closely connected with that of reparametrization and gauge
invariance of quantum gravity. Understanding the latter
circumstance may shed some new light on a possible solution of
this problem.

Our work consists of two parts. In the present part we shall remind
basic features of quantum geometrodynamics and minisuperspace
cosmological models, discuss fundamental problems of the Wheeler
-- DeWitt theory and give the outline of the paper (Section 2).
Further, in Section 3 we shall consider various attempts to find a
solution to the problem of time in the framework of canonical
approach. We shall investigate possible solutions making use of
minisuperspace models with finite degrees of freedom that will help
to clarify some points. However, in Section 4 we shall expand our
consideration beyond the minisuperspace approximation and review
briefly promising ideas by Brown and Kucha\v r \cite{ku95} and also
by Barvinsky and collaborators \cite{ba86a,ba86b,ba87,ba93}. The
program by Barvinsky, which can be presented both in the canonical
and in the path integral formalisms, is of great importance for
understanding the relationship between imposing a gauge condition
and introducing time in quantum gravity.

In the second part of our work we shall consider in more details
two approaches within the scope of Feynman path integration scheme.
The first approach by Simeone and collaborators
\cite{fesi97,si98,si99,si00,desi99b,gisi01a,gisi01b,gisi02,si02,si02b}
is essentially based on ideas by Barvinsky, in particular, on the
idea of deparametrization (reduction to physical degrees of
freedom). This proposal is gauge-invariant and lies in the course
of the unitary approach to quantization of gravity. The other
approach by Savchenko, Shestakova and Vereshkov
\cite{ssv99,ssv00,shest99,ssv01a,ssv01b} is rather radical. It is
an attempt to take into account peculiarities of the Universe as a
system without asymptotic states that leads to the conclusion that
quantum geometrodynamics constructed for such a system is, in
general, gauge-noninvariant theory. However, this theory is shown
to be mathematically consistent and the problem of time is solved
in this theory in a natural way.

\section{Quantum cosmology: basic issues}

\subsection{The gravitational field as a constrained system}

The Wheeler -- DeWitt quantum geometrodynamics is based upon
canonical quantization of constrained systems. The first step in
this procedure is rewriting of the Einstein -- Hilbert action $S$
which is a functional of the spacetime metric $g_{\mu\nu}(X)$ in
the Hamiltonian form. Then the dynamics is given by a succession of
spacelike three-dimensional hypersurfaces in four-dimensional
spacetime. By introducing the timelike parameter $\tau$ and the
internal coordinates $x^a$ ($a=1,2,3)$, the theory can be written
in terms of new set of variables: the spatial three-metric $g_{ab}$
on a hypersurface and the velocity $U^\mu$ with which this surface
evolves in spacetime. The normal and tangential components of the
velocity $U^\mu$ are the lapse and shift functions defined by
Kucha\v r \cite{ku76} as a generalization of those introduced by
Arnowitt, Deser and Misner \cite{adm63} $N=(-g^{00})^{-1/2}$,
$N^a=g^{ab}g_{b0}$. After the extrinsic curvature
$$
K_{ab}={1\over 2N}\left(\nabla_a N_b+\nabla_b N_a-{d g_{ab}\over d\tau}\right)
$$
describing the evolution of the spacelike hypersurface imbedded in
spacetime is defined, the Lagrangian form of the Einstein action
will be
\begin{eqnarray}
\lefteqn{S[g_{ab},N,N^a]= }\hspace{.7cm}\nonumber\\
& & \hspace{-1.5cm}\int_{\tau_1}^{\tau_2}  d\tau \int  d^3 x
\, N(
^3g)^{1/2} \left(K_{ab}K^{ab}-K^2+  ^3 R-2\Lambda\right)
\end{eqnarray}
where $^3 R$ is the scalar curvature of space, $K=g^{ab}K_{ab}$ and
$\Lambda$ is the cosmological constant.

The Hamiltonian form of the action is obtained defining the
canonical momenta
\begin{eqnarray}
\label{Gabcd}
p^{ab} & = & -2G^{abcd} K_{cd},\nonumber \\
G^{abcd} & = & {1\over 4}( ^3g)^{1/2}(g^{ac}g^{bd}+g^{ad}g^{bc}-2g^{ab}g^{cd}),
\end{eqnarray}
being $G^{abcd}$ the DeWitt supermetric. Then we have
\begin{eqnarray}
\label{grav.act}
\lefteqn{S[g_{ab}, p^{ab}, N, N^a]= }\nonumber\\
& &\int d\tau\int d^3x
 \left(p^{ab}{d g_{ab}\over d\tau}
-N{\cal H} -N^a{\cal H}_a\right),
\end{eqnarray}
where
\begin{eqnarray}
{\cal H} & = & {1\over 2}G_{abcd} p^{ab}p^{cd}-( ^3g)^{1/2}( ^3R-
2\Lambda),\nonumber\\
{\cal H}_a & = & -2 g_{ac}\nabla_d p^{cd},\nonumber\\
G_{abcd} & = & ( ^3g)^{-1/2} (g_{ac}g_{bd}+g_{ad}g_{bc}-2g_{ab}g_{cd}).
\end{eqnarray}

The lapse and shift functions are not determined; when we demand
the action to be stationary under an arbitrary variation of $N$ and
$N^a$ the {\it Hamiltonian and momentum constraints} are obtained:
\begin{equation}
\label{Ham.constr}
{\cal H}=0;
\end{equation}
\begin{equation}
\label{mom.constr}
{\cal H}_a=0.
\end{equation}
The presence of these constraints reflects the general covariance
of the theory. However, the status of the two constraints is
different: a basic role is given to the Hamiltonian constraint
(\ref{Ham.constr}) which generates dynamics of 3-geometry (the
change of canonical data under transition from some spacelike
hypersurface to another one). A dynamical character of the
Hamiltonian constraint results from non-standard quadratic
dependence of ${\cal H}$ from the momenta $p^{ab}$. It is the
reason why the Hamiltonian constraint has no analogy in other gauge
theories. The arbitrariness of $N$ leads to the so called
many-fingered nature of time: Because the lapse corresponds to the
velocity of the motion of the three-hypersurface in the normal
direction, as $N$ depends on $x^a$ and $\tau$ the separation
between two successive hypersurfaces is different in different
points of spacetime, and then the time has a local character.

The momentum constraints (\ref{mom.constr}) generate
diffeomorphisms of 3-metric $g_{ab}$ and are similar to constraints
in the Yang -- Mills theory. In their operator form after
quantization they are considered as the conditions that a wave
function is invariant under coordinate transformations of 3-metric.
Since the wave function is also independent on the lapse and shift
functions it leads to the conclusion that the wave function must
depend only on 3-geometry. But the latter statement remains to be
declarative: it has no mathematical realization. The wave
function always depends on a concrete form of the metric, which
gives rise to reparametrization noninvariance of the Wheeler --
DeWitt quantum geometrodynamics.

\subsection{Quantization and fundamental problems of the Wheeler -- DeWitt
theory}

In the Dirac canonical quantization the classical constraints are
turned into operators and are imposed on the wave function, which
must be annihilated by them. Hence the constraint ${\cal H}= 0$
leads to the Wheeler -- DeWitt equation
\begin{equation}
\label{WDWeq}
{\cal H}\Psi=0,
\end{equation}
A solution to this equation corresponding to observable physical
Universe is singled out by boundary conditions which acquire the
status of a fundamental law. However, this formulation of the
Wheeler -- DeWitt theory is not complete: such questions, as the
structure of Hilbert space or what quantities should be considered
as observables, remain open. At the same time these questions are
of great importance from the viewpoint of construction of any
quantum theory.

{\it The problem of time} is a consequence of the fact that the
gravitational Hamiltonian is a linear combination of constraints
(see (\ref{grav.act})) that leads to a static picture of the world.
DeWitt \cite{dew} commented it as following: Physical significance
can be ascribed only to intrinsic dynamics of the Universe while
its four-dimensional description, in particular, its evolution in
time, are irrelevant.

At the same time, any possible solution of {\it the problem of
Hilbert space} implies some solution of the problem of time. One
cannot determine the structure of Hilbert space if inner product of
state vectors is not defined. The inner product is to conserve in
time, so some definition of time is required. As a rule, time is
identified with a function of variables of configurational or
phase space. But in this case the status of time variable
differs from what it is in ordinary quantum mechanics, namely, an
extrinsic parameter related to an observer and marking changes in
a physical system.

Another problem which is closely connected with the problem of
time is {\it the problem of observables}. According to the Dirac
scheme, observables are quantities which have vanishing Poisson
brackets with constraints. It is indeed the true for
electrodynamics where all observables are gauge-invariant. But in
case of gravity this criterion leads to the conclusion that all
observables should not depend on time. Then one loses a
possibility to describe time evolution of a gravitational system
in terms of observables.

The next problem is that of {\it reparametrization noninvariance}:
At the classical level the gravitational constraints can be written
in various equivalent forms while at the quantum level, after
replacing the momenta by operators, these different forms of the
constraints become nonequivalent. It is a consequence of the fact
that the DeWitt supermetric $G^{abcd}$ depends, in general, on the
lapse function $N$ \cite{hp86} (in (\ref{Gabcd}) the choice $N=1$
has been made). In principle, one could replace $N$ with another
function of some new variable $\tilde N$ and 3-metric $g_{ab}$:
$N=v(\tilde N, g_{ab})$. This leads to changing the supermetric
$G^{abcd}$, so that corresponding Wheeler -- DeWitt equations would
have different solutions. A relation between these solutions can be
found in a very limited class of parametrizations \cite{hal88}. We
shall return to this point in Part II, where will be argued that
reparametrization noninvariance of the Wheeler -- DeWitt equation
can be understood as a hidden gauge noninvariance.

Let us also point to {\it the problem of global structure of
spacetime}. One can apply canonical quantization procedure only if
spacetime has the topology $R\times\Sigma$, where $\Sigma$ is some
3-manifold. In any other case it is impossible to introduce
globally (in the whole spacetime) a set of spacelike hypersurfaces
without intersections and other singularities, and it is
impossible to introduce a global time. In most papers simple
enough cosmological models are considered, and this problem seems
to be not so important. But the existence of this problem, as
well as previous ones, shows that the Wheeler -- DeWitt quantum
geometrodynamics needs to be modified.

\subsection{Interpretation}

Apart from mathematical problems, the Wheeler -- DeWitt quantum
geometrodynamics has no generally accepted interpretation. Of
course, the absence of a clear interpretation cannot be a reason
to revise a theory if the theory is mathematically consistent. In
the case of quantum geometrodynamics, however, the problem of its
interpretation results from its mathematical difficulties.

So, there does not exist {\it a precise probability interpretation}
of the wave function. It is related to the discussed above
mathematical problem of Hilbert space. Some authors have proposed
to start from a definition of time allowing to obtain a
Schr\"odinger equation \cite{ha86,cdf95,fe99,cafe01}; in this case
the physical inner product can be defined as
$$
(\Psi_2|\Psi_1)=\int dq\, \Psi_2^*\,{\hat\mu}\,\Psi_1,
$$
with ${\hat\mu}_{t'} =\delta (t-t')$, so that the integral is
evaluated at the fixed time $t'$. The central objection to such a
procedure is that the resulting wave functions are solutions of the
Wheeler--DeWitt equation (or can be related to them) only in the
case of a limited class of minisuperspace models.

Another possibility is to straightforwardly solve the
Wheeler--DeWitt equation in terms of a set of coordinates including
a global time \cite{vil86,ha86,gisi02,si02,si02b}:
$\{q^i\}=\{t,q^\gamma\}$; in this case the physical inner product
can be written
\begin{equation}
(\Psi_2|\Psi_1)={i\over 2}\int dq
 \left[\Psi^*_1{\partial\Psi_2\over\partial t}
 -\Psi_2{\partial \Psi^*_1\over\partial t}\right],
\end{equation}
where the integration is done at a fixed $t$ and is restricted to
the coordinates $q^\gamma$. This does not solve the problem of
defining a conserved positive probability, because a Klein--Gordon
inner product is obtained, which in general is not
positive-definite. Because the difficulties arise from the fact
that the Wheeler -- DeWitt equation is of second order in all its
derivatives, in the last years there have also been proposals based
on Dirac's solution to the problem, and some authors have
introduced a spinor wave function for cosmological models
\cite{mos98,mos0205,mos0209}.

Once adopting the Wheeler -- DeWitt theory, one should admit that a
wave function satisfying Eq.\,(\ref{WDWeq}) describes the past of
the Universe as well as its future with all observers being inside
the Universe in different stages of its evolution, and all
observations to be made by these observers. This picture might be
considered within the framework of the many-worlds interpretation
of the wave function proposed by Everett \cite{eve} and applied to
geometrodynamics by Wheeler \cite{wheel}. However, it does not seem
that the Wheeler -- DeWitt quantum geometrodynamics is a
mathematical realization of the Everett conception. Indeed, the
wave function satisfying Eq.\,(\ref{WDWeq}) and certain boundary
conditions is thought to be a branch of a many-worlds wave function
that corresponds to a certain universe; other branches being
selected by other boundary conditions. The information about all
possible actions of an observer through the whole history of the
Universe can be contained {\it only} in boundary conditions. At
the same time, any mathematical realization of the Everett
conception implies that a state of a closed system is a
superposition, each element of which is the product of some state
of the first subsystem and a relative state of the second one, one
of the subsystems being a measuring apparatus. To find such a
superposition for the Universe we need to define full sets of
orthonormal states of the subsystems that returns us to the
problem of Hilbert space.

Barvinsky and Ponomariov \cite{bapo86} discussed a mathematical
realization of the Everett conception. Though the full set of
orthonormal states was not defined, they showed that to define an
inner product in Hilbert space one should impose some gauge
condition that causes the wave function to depend in a certain
degree on this gauge condition. The state described by the wave
function was interpreted as a relative state of the Universe for
the chosen gauge condition. In Section 4.2 we shall comment the
central points of the unitary approach to quantum theory of gravity
proposed by Barvinsky. It is important that the work by Barvinsky
and Ponomariov has demonstrated that in any mathematical
realization of the Everett conception a wave function must contain
information about geometry of the Universe, as well as about a
reference frame, fixed by a gauge condition, in which this geometry
could be studied.

Further, the question arises if a theory, in which the wave
function depends on a gauge condition, could be gauge-invariant.
Barvinsky and his collaborators gave a positive answer to this
question. The quantization procedure proposed by them is thought to
be a "projection" of the gauge-independent Dirac
-- Wheeler -- DeWitt formalism \cite{ab96}. The equivalence with
the Dirac -- Wheeler -- DeWitt scheme can be proved in the one-loop
approximation and for some special quasiclassical states. In
\cite{ba88} Barvinsky wrote that the validity of extrapolating the
unitary approach to quantum cosmology is based on the success of
quantizing gauge theories in asymptotically-flat spacetime in
unitary gauges. We would note in this connection that the success
of quantization in asymptotically-flat spacetime is crucially based
on the presence of asymptotic states; the latter makes possible
to solve the full set of constraints and gauge conditions in the
limits of perturbation theory and split off the three-dimensionally
transversal gravitational degrees of freedom from the so-called
``nonphysical'' ones. In a general situation without asymptotic
states it may be that gauge invariance should be abandoned in a
formally consistent formulation. It is worth emphasizing that in
the both cases fixing a gauge enables one to introduce time in
quantum theory of gravity.

\subsection{The canonical approach and the outline of this paper}

The canonical approach in a wide sense unifies such methods as the
Dirac quantization \cite{di64} and the quantization in unitary
gauges which means a transition to a reduced phase space of true
physical degrees of freedom (see, for example, \cite{ba88}). These
methods are close to ordinary quantum theory in the sense that the
quantization procedure includes constructing a Hamiltonian
formalism.

This part of our work is entirely devoted to canonical methods. The
aim of Section 3 is to illustrate that without introducing a
physical time it is difficult to give a clear interpretation to
solutions to the Wheeler -- DeWitt equation for different models.
On the other hand, the minisuperspace approach, where one deals
with cosmological models with a finite number of degrees of
freedom, makes tractable searching for a wave function with all the
desired properties of a consistent theory, as a precise notion of
evolution and a well defined probability. We shall review most
representative developments within this line of work, namely, those
which start from different programs of deparametrization or
reduction to physical degrees of freedom as a previous step before
quantization \cite{ku81,ha86,ba86a,ba86b,ba87,
wa93,ba93,hiwa95,befe95,cdf95,kury97,cade97,caun99,fesf99,fe99}. In
Section 2.5 we shall remind main features of minisuperspace models
used in our below consideration.

A possible solution of the problem of time consists in the
identification of time with some function of variables of
configurational or phase space. To anticipate our consideration,
in Section 2.6 we shall formulate a condition to which admissible
functions must satisfy. The notion of an intrinsic and extrinsic
time will be also explained in this section.

The definition of time enables one to come to a Schr\"odinger
equation with a square-root true Hamiltonian. In Section 3.1 we
shall discuss a relation between the Schr\"odinger equation and the
Wheeler -- DeWitt equation and show that solutions to the
Schr\"odinger equation also satisfy the Wheeler -- DeWitt equation
if the Hamiltonian does not depend on the variable defined as time.

In Section 3.2 we shall demonstrate, following to H\'aj\'{\i}cek,
that the requirement of unitarity of a resulting theory may be
related to a right choice of a time variable. We shall touch upon
the WKB solutions to the Wheeler -- DeWitt equation in Section 3.3,
and it will be pointed out there that the definition of classically
forbidden and allowed regions is difficult for models where a clear
notion of time is absent.

The role of the identification of time will be illustrated in
Sections 3.4 -- 3.6 for the Taub universe. The behavior of the wave
function in minisuperspace leads to a certain choice for solutions
to the Wheeler -- DeWitt equation considered in Section 3.4, while
other solutions are discarded. On the other hand, the procedure of
identification of time based on the analogy with the ideal clock
results in the opposite choice for discarding the solutions.
Namely, as will be shown in Section 3.5, solutions to the
Schr\"odinger equation can be used to select a set of solutions to
the Wheeler -- DeWitt equation. In some cases it is possible to
define a phase time in such a way that the corresponding solution
to the Wheeler -- DeWitt equation would have an evolutionary form.
The example will be given in Section 3.6. In this situation we do
not need the Schr\"odinger equation to select solutions, though
there is a correspondence between the solutions to the Wheeler --
DeWitt equation with the time variable and those of the
Schr\"odinger equation considered in the preceding section. The
interpretation of these solutions is straightforward if the
Hamiltonian does not depend on time.

As was mentioned above, a possible way to introduce time into the
theory consists in imposing some gauge condition. In canonical
formalism it may be done by means of a time-dependent gauge
condition. This line of work will be discussed in Section 3.7. A
weak point of this approach is that different gauge choices lead
to non equivalent quantizations.

In Section 3.8 we shall consider a coordinate choice which gives
rise to the Wheeler -- DeWitt equation with a Hamiltonian not
depending on time. In this case there exists a direct
correspondence between the Wheeler -- DeWitt and Schr\"odinger
equations and their solutions. Unfortunately, this correspondence
exists only for a limited class of models.

Section 3.9 will be devoted to a rather exotic two-component
approach, in which the Wheeler -- DeWitt equation is reduced to a
system of first order equations with respect to time. It resembles
the transition from the Klein -- Gordon equation to the Dirac one
and requires introducing a spinor wave function. This procedure
also leads to a Schr\"odinger equation and an appropriate
interpretation.

The disadvantage of the methods presented in Section 3 is that they
can be applied to restricted classes of models. Their application
depends in a large degree on the choice of suitable coordinates and
a resulting form of the Hamiltonian constraint. In Section 4 we
shall review briefly general approaches formulated for the full
gravitational theory. In Section 4.1 we shall consider an
interesting idea by Brown and Kucha\v r \cite{ku95} that dust
interacting only gravitationally can serve as a time variable. This
proposal leads to a special form of the constraints and,
eventually, to a Schr\"odinger equation. It lies entirely within
the scope of the canonical formalism. On the other hand, the
approach by Barvinsky \cite{ba86a,ba86b,ba87,ba93}, which we have
already mentioned in Section 2.3 and whose main points we shall
remind in Section 4.2, admits both the canonical and path integral
formulations. It anticipates the consideration of recent
developments in the path integral approach in the second part of
our work.

\subsection{The minisuperspace models}

If all except a finite number of degrees of freedom of the
classical theory are set to zero, we obtain the {\it
minisuperspace approximation}; the choice of an homogeneous lapse
and zero shift lead to an action whose Hamiltonian form is
\begin{equation}
S[q^i,p_i,N]=\int_{\tau_1}^{\tau_2} \left( p_i{dq^i\over
d\tau}-N{\cal H}\right)d\tau,
\label{28}
\end{equation}
where
\begin{equation}
{\cal H}=G^{ij}\,p_ip_j+V(q).
\end{equation}
Here $G^{ij}$ is the reduced version of the DeWitt supermetric and
$V$ is the potential, which depends on the curvature and includes
terms corresponding to the coupling between the gravitational field
and matter fields; it is understood that a spatial integration has
already been performed, so that only the integration on $\tau$
remains. As the shift is null the momenta read
$p_i={1\over N}G_{ij}{dq^j\over d\tau}$.
On the classical path  we have  the Hamilton canonical equations
\begin{equation}
{dq^i\over d\tau}=N[q^i,{\cal H}],\quad
{dp_i\over d\tau}=N[p_i,{\cal H}]
\end{equation}
and the minisuperspace version of the Hamiltonian constraint
$${\cal H}=0.$$
The evolution of the lapse $N$ is arbitrary, as it is not
determined by the canonical equations. Hence, the separation
between two successive spatial three-surfaces, although globally the
same, is still undetermined: this is the minisuperspace version of
the many-fingered nature of time of the full theory.

The spatial line element of an isotropic and homogeneous
cosmological model has the form
$$
dl^2=g_{ab}dx^adx^b
$$
where $g_{ab}$ is the space metric, whose components are functions
of time. The isotropy and homogeneity hypothesis lead to the fact
that the curvature depends on only one parameter: for $k=0$ we have
a flat universe, for $k=-1$ the universe is open, and for $k=1$ the
universe is closed. The spacetime metric has then the
Friedmann -- Robertson -- Walker form \cite{dau75}
\begin{eqnarray}
ds^2 & = &
 N^2d\tau^2\nonumber\\
& & \hspace{-1cm}\mbox{} -a^2(\tau)\left({dr^2\over
1-kr^2}+r^2d\theta^2+r^2\sin^2\theta d\varphi^2\right),
\end{eqnarray}
where $a(\tau)$ is the spatial scale factor.

The hypothesis of homogeneity and isotropy completely determines
the form of the space metric leaving free only the curvature;
restricting the hypothesis to homogeneity without any other
symmetry assumption allows for much more freedom. Homogeneity
implies that the metric properties are the same at any point of
space. The mathematical formulation of this is given by the set of
transformations which leave the metric unchanged. For an
homogeneous non-Euclidean space, the transformations of the
symmetry group leave invariant three linear differential forms;
these forms are not total differentials of functions of the
coordinates, but they read
$$
\sigma^i=e^i_a dx^a
$$
where $a=1,2,3$ and $e^i$ are three independent vectors. The
differential forms fulfill
$d\sigma^i=\epsilon_{ijk}\sigma^j\times\sigma^k$. The invariant
space metric can then be written as \cite{dau75}
$$
dl^2=g_{ij}\sigma^i\sigma^j=g_{ij}(e^i_a dx^a)(e^j_b dx^b),
$$
so that the spatial metric tensor has the components
$$
g_{ab}=g_{ij}e^i_a e^j_b.
$$
Possible anisotropic cosmologies are comprised by the Bianchi
models and the Kantowski -- Sachs model \cite{rysh75}. By introducing
the diagonal $3\times 3$ matrix $\beta_{ij}$ their spacetime
metrics can be written as
\begin{equation}
ds^2=
N^2d\tau^2-e^{2\Omega(\tau)}(e^{2\beta(\tau)})_{ij}\sigma^i\sigma^j.
\label{450}
\end{equation}
However, the spatial geometry of Bianchi models is essentially
different from that of the Kantowski -- Sachs model, because a
continuous transformation carrying from the last one to the Bianchi
form does not exist.

\subsection{Global phase time}

A globally good time is a function $t(q^i,p_i)$ which monotonically
increases along a dynamical trajectory, that is, each surface
$t=\rm const$ in the phase space is crossed by a dynamical
trajectory only once; hence the successive states of the system can
be parametrized by this function. This means that $t(q^i,p_i)$
must fulfill the condition
\be
{\rm H}^A{\partial t\over \partial x^A}>  0
\ee
where ${\rm H}^A$ are the components of   the Hamiltonian vector
\be
{\bf H} \equiv  ({\rm H}^q,{\rm H}^p)
=  \left({\partial {\cal H}\over \partial p},-{\partial {\cal H}\over\partial
q}\right).
\ee
The definition of Poisson brackets clearly leads to the equivalent
condition \cite{ha86}
\be
[t,{\cal H}]>  0.
\ee
(If we  define a scaled constraint
$$
H={\cal F}^{-1}{\cal H},\quad
{\cal F}> 0,
$$
it can easily be shown that $H$ and ${\cal H}$ are equivalent, in
the sense that they describe the same parameterized system: their
field lines, which coincide with the classical trajectories, are
proportional on the constraint surface. Thus, if we can find a
function $\overline t(q^i,p_i)$ with the property
$$
[\overline t,H]>  0,
$$
we know that $\overline t(q^i,p_i)$ monotonically increases along
the dynamical trajectories associated to both $H$ and ${\cal H}$,
and it is also a global phase time.)

Because the supermetric $G^{ik}$ does not depend on the momenta, a
function $t(q^i)$ is a global time if the bracket
\be
[t(q^i),{\cal H}] =  [t(q^i), G^{ik}p_ip_k]
 = 2{\partial t\over\partial q^i} G^{ik}p_k
 \ee
is positive definite. Hence if the supermetric has a diagonal form
and one of the momenta vanishes at a given point of phase space,
then no function of only its conjugated coordinate can be a global
time. For a constraint whose potential can be zero for finite
values of the coordinates, the momenta $p_k$ can be all equal to
zero at a given point, and $[t(q^i),{\cal H}]$ can vanish. Hence an
{\it intrinsic time} $t(q^i)$ \cite{ku92} exists only if the potential in the
constraint has a definite sign. In the most general case a global
phase time should be a function including the canonical momenta;
this is called an {\it extrinsic time} $t(q^i,p_i)$ \cite{ku71,yo72}, because the
momenta are related to the extrinsic curvature $K_{ab}$  which
describes the evolution of spacelike three-dimensional
hypersurfaces in four-dimensional spacetime: in the case of no matter fields, we have
$$
p_i\equiv p^{ab}= -2G^{abcd} K_{cd}.
$$
 The existence of a time in
terms of only the coordinates is related to the fact that, in some
special cases which do not represent the general features of
gravitation, the coordinates can be obtained in terms of the
momenta with no ambiguities; however, this is not always possible,
and a consistent quantization can require to work with an extrinsic
time.

\section{Canonical quantization}

Imposing the operator form of the original Hamiltonian constraint
on a wave function yields the usual Wheeler--DeWitt equation, which
is of second order in all its derivatives. A Schr\"odinger
equation, instead, requires a previous definition of time, and then
it includes the notion of a true (non-vanishing) Hamiltonian.
Though the Wheeler--DeWitt equation is the most common choice for
the canonical quantization of minisuperspaces, it makes difficult
the interpretation of the resulting wave function in terms of a
conserved positive-definite inner product. The Schr\"odinger
quantization, instead, allows to define a conserved inner product,
and then a clear probability interpretation can be given for the
wave function.

In some cases, a Schr\"odinger equation has been obtained by
splitting the constraint into two disjoint sheets given by the two
signs of the momentum $p_0$ conjugated to a coordinate $q^0$
identified as time; this yields a canonical quantization consisting
in two equations of first order in $\partial/\partial q^0$. Thus we
have a pair of Hilbert spaces, each one with its corresponding
Schr\"odinger equation. In this case we can say that the
Schr\"odinger quantization preserves the topology of the constraint
surface, that is, the splitting of the classical solutions into two
disjoint subsets has its quantum version in the splitting of the
theory into two Hilbert spaces \cite{si03}.

The subtleties involved in the splitting of the original constraint
into two constraints, namely $K^+=0$ and $K^-=0$, were first
carefully considered by Blyth and Isham \cite{isham}. These two
constraints together are classically equivalent to the original
Hamiltonian constraint $H=0$, which is quadratic in all the
momenta; that is, classical dynamics take place in one of the two
sheets of the constraint surface determined by the sign of a
non-vanishing momentum. But at the quantum level this equivalence
is no more fulfilled if time appears in the potential: a function
in the kernel of the operator $\hat K^+$ or $\hat K^-$ is not
annihilated by the operator $\hat H$, but by $\hat H$ plus terms
corresponding to a commutator between $\hat p_0$ and the
square-root true Hamiltonian resulting from its time-depending
potential. It must be emphasized that these terms cannot be
eliminated by operator ordering.

We shall then begin with a discussion of the mentioned work by
Blyth and Isham; also, the thorough discussion by H\'aj\'{\i}cek
about the relation between unitarity and the identification of time
\cite{ha86} is reproduced and commented in detail. Then we shall
follow with a review of two standard procedures, one by Halliwell
and the other one by Moncrief and Ryan, in the framework of a
Wheeler--DeWitt equation straightforwardly obtained from the
constraints of different homogeneous models {\it without} a
previous analysis about the problem of time. We shall emphasize the
unsatisfactory points of such procedure, and then show some
improvements based on the identification of a global time as a step
before quantization. Within this context, we shall also discuss the
role of a Schr\"odinger equation, both as an auxiliary tool for
selecting solutions of the Wheeler-DeWitt equation, and as the
central equation for quantization; in particular, we analyse the
case of an interesting approach starting with the identification of
time by means of gauge fixation, and also a two component
formulation inspired in Dirac's solution to the problems of the
Klein--Gordon equation.

\subsection{Wheeler--DeWitt equation and  Schr\"odinger equation}

A good introduction to the problem of choosing between these two
formulations can be found in the early work by Blyth and Isham,
Ref. \cite{isham}. Within the context of the canonical quantization
of a Friedmann--Robertson--Walker universe with matter in the form
of a scalar field, the authors carefully study the reduction
procedure leading to a Schr\"odinger equation and establish its
inequivalence with the standard Wheeler--DeWitt approach. The
analysis starts with the identification of one of the canonical
coordinates of the model as time variable (in practice, the scale
factor; see below), thus reducing the system and treating it in the
usual canonical form with a true time and a true non-vanishing
Hamiltonian. The result is a time-dependent Schr\"odinger equation
\be
i{\partial\Psi\over \partial t}=h\Psi,\label{sch}
\ee
where $h$ is a square-root true Hamiltonian. This requires a
definition by means of the spectral theorem, assuming that the
square root is taken on a positive definite self-adjoint operator.
This point relies on the right identification of time (see the next
Section); for example, the scale factor is a bad time variable for
any model allowing for $p_\Omega =0$.

The usual Wheeler--DeWitt approach would lead, instead of
(\ref{sch}), to the second order equation
\be
-{\partial^2\Psi\over \partial t^2}=h^2\Psi,\label{wdw}
\ee
which in the most general case is not equivalent to (\ref{sch}). In
fact, by acting with $h$ on both sides of the Schr\"odinger
equation the result obtained is
 \be
-{\partial^2\Psi\over \partial t^2}-i{\partial h\over\partial
t}\Psi=h^2\Psi.\label{xx}
\ee
Clearly the solutions of (\ref{sch}) and (\ref{wdw}) will then be
different, unless the potential in the square-root Hamiltonian $h$
does not depend on the variable defined as time. In the case that
$h$ commutes at different times, the integration of (\ref{sch})
yields
\be
\Psi(x,t)=\exp\left(-i\int_{t_0}^th(s)ds\right)\Psi(x,t_0),
\ee
where $x$ stands for the true degrees of freedom of the system. A
decomposition in terms of eigenstates $\Psi_E(x,t)$ can be given,
with
$$
\Psi_E(x,t)=\exp\left(-i\int_{t_0}^tE(s)ds\right)\Psi_E(x,t_0),
$$
and $\Psi_E(x,t_0)$ a solution of the equation
$$
h^2(x,t_0)\Psi_E(x,t_0)=E^2\Psi_E(x,t_0).
$$
Here there is no problem with the square of the true Hamiltonian
$h^2$ because this is an eigenvalue equation for a fixed time
$t_0$. Once a definite solution is obtained, then it can be
provided with physical meaning because the corresponding inner
product is well defined, which is not the case for the
Klein--Gordon type equation (\ref{wdw}). Though the first choice of
time by the authors is the scale factor, also other choices are
explored, including extrinsic times. This is unavoidable for any
Friedmann--Robertson--Walker model with a constraint including a
potential which can be zero for finite values of the coordinates.

\subsection{Unitarity and time}

Here we shall reproduce and analyse the early work by
H\'aj\'{\i}cek where the relation existing between a right choice
of time and the obtention of a unitary theory \cite{ha86} is
clearly established, and the analogy between the existence of a
global phase time for a parametrized system and the possibility a
globally good gauge choice for a gauge system is discussed.

Instead of the models studied by H\'aj\'{\i}cek, we shall consider
a generic (scaled) constraint of the form
\begin{equation}
-\tilde p_1^2+\tilde p_2^2+Ae^{(a\tilde q^1+b\tilde q^2)}=0\label{613}
\end{equation}
with $a\neq b$, and where we have used tildes to denote that the
variables are not necessarily the original ones, but a set
$\{\tilde q^i,\tilde p_i\}$ including the coordinate $\tilde q^0$ which is a
global time. This
Hamiltonian corresponds to some models of interest, like dilaton
cosmologies, the Kantowski--Sachs universe and even the Taub
universe after a canonical transformation. It is easy to show that
a coordinate change exists leading to
\begin{equation}
H=-p_x^2+p_y^2+\zeta e^{2x}= 0
\end{equation}
with $sgn (\zeta)=sgn (A/(a^2-b^2))$. Depending on the sign of the
constant $A$ in the constraint (\ref{613}), these models admit as
global phase time the coordinates $x$ or $y$. In the case $\zeta>0$
the time is $t=\pm x$, so that following Ref. \cite{ha86} we can
define the reduced Hamiltonians as $h_{\pm}=\pm\sqrt{p_y^2+\zeta
e^{2x}}$, and we can write the Schr\"odinger equations
\begin{equation}
i{\partial\over\partial x}\Psi(x,y)=\mp\left(-{\partial^2\over\partial
y^2}+\zeta e^{2x}\right)^{1/2}\Psi(x,y)
\end{equation}
(note that in this case we obtain a time-dependent potential). If,
instead, we have $\zeta<0$, the time is $t=\pm y$ and the reduced
Hamiltonians corresponding to each sheet of the constraint surface
are $h_{\pm}=\pm\sqrt{p_x^2-\zeta e^{2x}}$; the associated
Schr\"odinger equations are
\begin{equation}
i{\partial\over\partial y}\Psi(x,y)=\mp\left(-{\partial^2\over\partial x^2}-
\zeta e^{2x}\right)^{1/2}\Psi(x,y).
\end{equation}
For both $\zeta>0$ and $\zeta<0$ we have a pair of Hilbert spaces,
each one with its corresponding Schr\"odinger equation, and a
conserved positive-definite inner product allowing for the usual
probability interpretation for the wave function. This is analogous
to the obtention of two quantum propagators, one for each disjoint
theory, mentioned in the context of path integral quantization
\cite{si03,ba93}.

The point to be remarked is that as a result of the right time
definition, in both cases the reduced Hamiltonians are real, so
that the evolution operator is self-adjoint and the resulting
quantization is unitary. Instead, a wrong choice of time, like for
example $t=\pm x$ in the case $\zeta<0$, leads to a Hamiltonian for
the reduced system which is not real for all allowed values of the
variables, and we obtain a nonunitary theory.

There is an aspect, however, which should be signaled, though it
was not emphasized in Ref. \cite{ha86}. In the case $\zeta>0$ the
Schr\"odinger equations can be obtained by starting from the
constraint written as a product of two linear constraints
 \begin{equation}
\left(-p_x+\sqrt{ p_y^2+\zeta e^{2x}}\right)\left( p_x+\sqrt{ p_y^2+\zeta
e^{2x}}\right)= 0,
\end{equation}
and it is then clear that the potential depends on time. Therefore
though at the classical level this product is equivalent to the
constraint (\ref{613}), in its operator version both constraints
differ in terms associated to commutators between  $ p_x$ and the
potential $\zeta e^{2 x}$. Hence, depending on which of the two
classically equivalent constraints we start from, we obtain
different quantum theories. Observe that this problem appears in
the case for which the Wheeler--DeWitt equation leads to a result
in which the identification of positive and negative-energy
solutions is not apparent, at least in the standard form: for the
case $\zeta >0$, $t=\pm x$ we obtain the Wheeler--DeWitt solutions
\begin{eqnarray}
\Psi _\omega (x,y) & = & \left[ a_{+}(\omega )e^{i\omega y}+a_{-}(\omega )e^{-
i\omega y}\right]\nonumber\\
& &\hspace{-1.7cm}
\times \left[ b_{+}(\omega )J_{i\omega }(\sqrt{\left| \zeta \right| }e^x)+b_{-
}(\omega )N_{i\omega
}(\sqrt{\left| \zeta \right| }e^x)\right],
\end{eqnarray}
with $J_{i\omega} $ and $N_{i\omega}$ the Bessel and Neumann
functions of imaginary order respectively; note that the time
dependence appears in the argument of Bessel functions. Instead,
for $\zeta <0$,  $t=\pm y$, the solutions are of the form
\begin{eqnarray}
\Psi _\omega (x,y) & = & \left[ a_{+}(\omega )e^{i\omega y}+a_{-}(\omega )e^{-
i\omega y}\right]\nonumber\\
& &\hspace{-1.7cm} \times\left[ b_{+}(\omega )I_{i\omega }(\sqrt{\left| \zeta \right|
}e^x)+b_{-}(\omega )K_{i\omega
}(\sqrt{\left| \zeta \right| }e^x)\right],  \label{619}
\end{eqnarray}
where $I_{i\omega}$ and $K_{i\omega}$ are modified Bessel
functions. In this case the usual factors $\sim e^{i\omega t}$
associated to definite-energy states are obtained; moreover, now
the Wheeler--DeWitt solutions are the same corresponding to  the
Schr\"odinger equation, so that the inner product is well defined.

We insist on the point regarding the topology of the constraint:
The choice of the Schr\"odinger formulation always preserves the
classical geometry of the constraint surface \cite{car94,si03}; in
the case of a time-dependent potential this is achieved by
introducing the commutator mentioned above, whose  form
is $\left[\sqrt{\sum(\hat{\tilde p_r})^2 +V(\hat{\tilde
q^i})},\hat{\tilde p}_0\right]$ (where $r\neq 0$, and $V$ stands
for the potential in the scaled Hamiltonian constraint $H$). It is
clear that this cannot be avoided by any operator ordering.

\subsection{Approximate solutions of the Wheeler--DeWitt equation}

The impossibility of explicitly integrating the Wheeler--DeWitt
equation except for a limited class of models has led to several
attempts of quantization based on approximations valid for
different regions of the phase space. Consider the Hamiltonian
constraint of a closed $(k=1)$ homogeneous and isotropic universe
with a scalar field $\phi$ and null cosmological constant; assume a
generic dependence of the potential with $\phi$, namely $V(\phi)$.
The associated Wheeler--DeWitt equation obtained by replacing $p
\to -i\partial/\partial q$ (and considering the trivial factor
ordering) reads
\begin{equation}
\left({\partial^2\over\partial\Omega^2}-
{\partial^2\over\partial\phi^2}+V(\phi)e^{6\Omega}-
e^{4\Omega}\right)\Psi(\Omega,\phi)=0.
\end{equation}
Halliwell has analysed the region of phase space such that $\vert
V'/V\vert \ll 1$ and found solutions whose variation with the
matter field is small, so that the $\phi$ derivative can be
neglected. In the region where the scale factor is small the
resulting WKB solutions have the exponential form \cite{hal90}
\begin{equation}
\Psi(\Omega,\phi)\sim \exp\left(\pm{1\over 3V(\phi)}(1-
e^{2\Omega}V(\phi))^{3/2}\right),
 \end{equation}
and are associated to a classically forbidden region. For large
values of the scale factor the WKB solutions have the oscillatory
form
\begin{equation}
\Psi(\Omega,\phi)\sim \exp\left(\pm{i\over 3V(\phi)}(e^{2\Omega}V(\phi)-
1)^{3/2}\right).
 \end{equation}
These solutions correspond to what is considered the classically
allowed region. Both kinds of solution can be matched by means of
the usual WKB matching procedure. In the case
$e^{2\Omega}V(\phi)\ll 1$ it can be shown that the oscillatory wave
function is peaked about a solution of the form
$$e^\Omega\sim e^{{\sqrt V}\tau},~~~~~~~~~~~~~\phi\sim\phi_0,$$
which corresponds to an inflationary behaviour. (For the case
$V(\phi)$=0 an exact solution can be easily obtained in terms of
modified Bessel functions. This is also the case if $V(\phi)$=0 in
a flat $(k=0)$ model with nonzero cosmological constant). Depending
on the form of $V(\phi)$, the regions considered by Halliwell may
be related to those to which the analysis should be restricted if
one was to define an intrinsic time in the case of models for which
this cannot be done globally. We should signal that the absence of
a notion of time within this formulation, besides making not clear
the interpretation of the formalism, makes not completely justified
the identification of classically forbidden or allowed regions, as
this would require a separation between true degrees of freedom and
time; we return to this point, with more detail, in the following
paragraph.

\subsection{Exact solutions without time}

In the literature we can find different exact solutions for the
Wheeler--DeWitt equation for minisuperspace models. An example
among those which do not start from an explicit deparametrization
is the solution found for the Taub universe (see the next section) by Moncrief
and Ryan
\cite{mory91} in the context of an analysis of the Bianchi type-IX
universe with a rather general operator ordering of the Hamiltonian
constraint \cite{haha83}. In the case of the most trivial ordering
they found the following general solution for the Wheeler -- DeWitt
equation:
\begin{eqnarray}
\Psi(\Omega,\beta_+)& = & \int_0^\infty d\omega F(\omega)\nonumber\\
& &\hspace{-1.2cm}\mbox{}\times  K_{i\omega}\left({1\over 6}
e^{2\Omega-4\beta_+}\right) K_{2i\omega}\left({2\over 3} e^{2\Omega-
\beta_+}\right),
\end{eqnarray}
where $K_{i\omega}$ are modified Bessel functions of imaginary
argument; modified functions $I$ would also appear, but they are
discarded because they are not well behaved for $\beta_+\to
\pm\infty$ (see below). In the particular case that $F(\omega)=
\omega \sinh (\pi\omega)$ Moncrief and Ryan showed that the wave
function can be written in the form
\begin{equation}
\Psi(\Omega,\beta_+)=R(\Omega,\beta_+) e^{-S}
\end{equation}
with
$$S={1\over 6}e^{2\Omega}\left(e^{-4\beta_+} +2e^{2\beta_+}\right).$$
This wave function has the nice feature that for values of $\Omega$
near the singularity (that is, the scale factor near to zero) the
probability is spread over all possible degrees of anisotropy given
by $\beta_+$, while for large values of the scale factor the
probability is peaked around the isotropic
Friedmann--Robertson--Walker closed universe; the authors, however,
prevent from a naive interpretation of the wave function, and they
note that there are different probability interpretations that
would not agree with this one.

There are two central objections to this straightforward procedure,
and both arise from the absence of a notion of time in the
formalism: first, because time has not been identified it is not
possible to speak about a conserved probability, hence the meaning
of the wave function is not at all clear. Second, the choice of a
set of solutions made by discarding the modified Bessel functions
$I$ would only be justified by the bad behaviour of the wave
function in a region of the configuration space defined by the
form of the potential of {\it a true Hamiltonian}. This is not
the case, because a true Hamiltonian is necessarily related to a
physical time, which is lacking in this formulation; in fact, we
shall immediately see that a careful analysis of this point leads
to exactly the opposite choice for discarding Bessel functions. The
procedure (see the next section) will be based in an intermediate
line of work consisting in combining the Wheeler--DeWitt equation
with a Schr\"odinger equation.

\subsection{Boundary conditions for Wheeler--DeWitt solutions  from a
Schr\"odinger equation}

The problem mentioned in the two preceding examples could be solved
by an approach beginning with the identification of a global phase
time like that in Ref. \cite{cafe01}, whose authors obtain a
Schr\"odinger equation and they use its solutions to select a set
of solutions of the Wheeler--DeWitt equation. The underlying idea
is that the typical constraint of a parametrized system, which is
linear in the momentum conjugated to the true time, is hidden in
the formalism of gravitation. This is an extension of the analogy
between the ideal clock and empty isotropic models
\cite{befe95,fe99,desi99a}:
The constraint of the ideal clock
$${\cal H}=p_t-t^2= 0$$
yields the Schr\"odinger equation
\begin{equation}
i{\partial\Psi\over\partial t}= -t^2\Psi\label{clock}
\end{equation}
which is of parabolic form, and it has the only solution
$\Psi=e^{it^3/3}.$ As a first step to obtain the constraint of a
minisuperspace, a canonical transformation leading to a constraint
quadratic in the momenta is performed: defining $Q=p_t$, $P=-t$, we
obtain
$${\cal H}=-P^2+Q= 0.$$
(The Hamiltonian of empty isotropic models results from the second
transformation $Q=\tilde V(\Omega)$, $P=p_\Omega(d\tilde
V/d\Omega)^{-1}$, with $\tilde V$ the potential defined in Ref.
\cite{desi99a}). The differential equation associated to the
constraint is now of hyperbolic form:
\begin{equation}
{\partial^2\Psi\over\partial Q^2}+Q\Psi=0.
\end{equation}
As this equation is of second order, it has two independent
solutions, which are the Airy functions $Ai(-Q)$ and $Bi(-Q)$. The
central point is that while $Bi(-Q)$ diverges for $Q\to-\infty$,
$Ai(-Q)$ is well behaved (in fact, it vanishes) in this limit, and
it is the Fourier transform of the solution of Eq. (\ref{clock}).
This provides a rule for selecting solutions of the hyperbolic
equation: the physical solutions are those which are in
correspondence with the solutions of the Schr\"odinger equation.

This line is then followed in \cite{cafe01} for quantizing the Taub
universe, which is a particular case of the Bianchi type IX model
\cite{dau75,rysh75}. In the absence of matter, its Hamiltonian
constraint reads
\begin{equation}
H=p_+^2-p_\Omega^2+{1\over 3}e^{4\Omega} (e^{-8\beta_+}-4e^{-2\beta_+})= 0,
\end{equation}
where $\beta_+$ determines the degree of anisotropy. The Taub
universe involves a potential which vanishes for finite values of
the coordinates, so making impossible the definition of an
intrinsic time in terms of the original variables. By defining
$
x=\Omega-2\beta_+$, $ y=2\Omega-\beta_+
$
the constraint can be put in the form
\begin{equation}
H= p_x^2-p_y^2+{1\over 9}(e^{4x} -4e^{2y})=0.\label{469}
\end{equation}
(the authors work with a different choice of the constants); then
the corresponding Wheeler--DeWitt equation
\begin{equation}
\left({\partial^2\over\partial x^2}- {\partial^2\over\partial y^2}-{1\over
9}e^{4x}+{4\over 9}e^{2y}\right)\Psi(x,y)=0
\end{equation}
is solved as it was done by Moncrief and Ryan. The authors obtain
the solutions
\begin{eqnarray}
\Psi_\omega(x,y) & = & \left[a(\omega)I_{i\omega}\left({2\over
3}e^y\right)+b(\omega)K_{i\omega}\left({2\over 3}e^y\right)\right]\nonumber\\
& & \hspace{-1.8cm}\times\left[c(\omega)I_{i\omega/2}\left({1\over
6}e^{2x}\right)+d(\omega)K_{i\omega/2}\left({1\over 6}e^{2x}\right)\right],
\end{eqnarray}
with $I$ and $K$ modified Bessel functions. Then they consider a
canonical transformation generated by
\begin{equation}
\Phi_1 (y,s)= -{2\over 3}e^y \sinh s
\end{equation}
leading to the following form for the  Hamiltonian constraint:
\begin{equation}
H(s,x,\pi_s,\pi_x)=-p_s^2+ p_x^2+{1\over 9}e^{4x}=0\label{472}
\end{equation}
so that the momentum $p_s$ is negative definite, and the time is
$t=s$; hence the constraint is written as a product of two factors
linear in $p_s$, the first one positive definite, and the second
one a constraint including a true Hamiltonian
$h=\sqrt{p_x^2+(1/9)e^{4x}}$ which does not depend on time (as we
have already remarked, this makes possible the equivalence of the
linear constraint and the original quadratic one). This constraint
then leads to the Schr\"odinger equation
\begin{equation}
i{\partial\over\partial t}\Psi(x,t)= \left(-{\partial^2\over\partial
x^2}+{1\over 9}e^{4x}\right)^{1/2}\Psi(x,t).
\end{equation}
It is necessary a prescription to give a precise meaning to the
Hamiltonian operator; the square root containing the derivative
operator must be understood as its binomial expansion, which allows
to propose solutions of the form $\sim \phi(x)e^{-i\omega t}$.
According to this interpretation, the contribution of the functions
$I_{i\omega/2}((1/6)e^{2x})$ is discarded, because they diverge in
the classically forbidden region associated to the exponential
potential ${1\over 9}e^{4x}$; the functions
$I_{i\omega}((2/3)e^y)$, instead, are not discarded, because in
this picture the coordinate $y$ is associated to the definition of
time. In fact, by transforming the solutions of the Wheeler--DeWitt
equation it is shown that those corresponding to the solutions of
the Schr\"odinger equation are precisely the functions
$I_{i\omega}((2/3)e^y)$, while the functions
$K_{i\omega}((2/3)e^y)$ must be ruled out because they cannot be
associated to definite energy states of the true Hamiltonian $h$.
It is remarkable that the functions in the selected subspace do not
decay in what was previously considered a classically forbidden
zone; note then the central difference with the result of the
preceding paragraph.

\subsection{Wheeler--DeWitt equation with extrinsic time}

A possible deparametrization and canonical quantization program can
be to start from a form of the Hamiltonian constraint such that a
global phase time is easily identified as one of the canonical
coordinates (in the general case, this could require a previous
canonical transformation); this is reflected in the corresponding
Wheeler--DeWitt equation, and hence the resulting wave function has
an evolutionary form. If the  reduced Hamiltonian does not
depend on time, the wave function then may be interpreted as it is
in ordinary quantum mechanics. We shall illustrate this line of
work with a solution for the Taub universe \cite{gisi02}.

If we admit a double sign in the generating function for the
canonical transformation leading to a constraint with a non
vanishing potential, then the Hamiltonian (\ref{472}) allows to
immediately define the time as
$$t= -s\, sgn (p_s).$$
As we shall see in Part II, this time can be obtained by choosing
a simple canonical gauge condition, which in the variables $\tilde
q^i$ including a global time has the form $s=\eta T(\tau)$,
$\eta=\pm 1$. The corresponding Wheeler--DeWitt equation is
\begin{equation}
\left({\partial^2\over\partial x^2}- {\partial^2\over\partial s^2}-{1\over
9}e^{4x}\right)\Psi(x,s)=0.
\end{equation}
This  equation has the set of solutions \cite{gisi02}
\begin{eqnarray}
\Psi_{\omega}(x,s) & = & \left[a(\omega)e^{i\omega s}+b(\omega)e^{-i\omega
s}\right]\nonumber\\
& & \hspace{-1.8cm}\times\left[c(\omega)I_{i\omega/2}\left({1\over
6}e^{2x}\right)+d(\omega)K_{i\omega/2}\left({1\over 6}e^{2x}\right)\right],
\label{617}\end{eqnarray}
where $\pm s$ is a global phase time. The contribution of the
functions $I_{i\omega/2}$ should be discarded as they are not well
behaved for great values of $x$; now this is completely justified,
as the exponential $\sim e^{4x}$ is the potential of a true
Hamiltonian. Then by recalling that $t=\pm s$ the wave function can
be given in terms of a set of definite-energy solutions:
\begin{equation}
\Psi_{\omega}(x,t)=a(\omega)e^{-i\omega t}K_{i\omega/2}\left({1\over
6}e^{2x}\right).
\end{equation}
This reflects that both theories corresponding to both sheets (in
terms of the new variables) of the constraint surface are
equivalent \cite{si02}.

The solutions of this Wheeler--DeWitt equation correspond to those
of the Schr\"odinger equation of the preceding section. This
procedure allows to obtain them without the necessity of defining a
prescription for the square root operator, but only by choosing the
trivial factor ordering; differing from the previous treatment, now
these solutions are not merely considered as a tool for imposing
boundary conditions, but are understood as the wave function for
the model. A point to be remarked is that in this description the
role of the original momenta, though unavoidable provided the
topology of the constraint surface in the original variables, is
restricted to the global phase time
$s=\pm\mbox{arcsinh}\left({1\over 2}(p_\Omega+p_+)
e^{(-2\Omega+\beta_+)}\right)$; the other coordinate entering the
wave function is a simple function of only the original
coordinates.

Though this procedure is the most straightforward including a right
notion of time, an unsatisfactory point is that the resulting wave
function can be interpreted in terms of probabilities because the
constraint of the model considered here leads to the same solutions
for both the Wheeler--DeWitt and the Schr\"odinger equations; hence
we can define the probability by means of the ordinary
Schr\"odinger inner product, which is conserved and
positive-definite. In the case of a model with a time-depending
potential this would not be possible, and though we could isolate
time and obtain an evolutionary wave function, its meaning would be
not at all clear (see, however, Section 3.8 for a possible solution
for a limited class of models).

\subsection{Gauge fixation and Schr\"odinger equation for isotropic models}

As we have already pointed, the close relation existing between the
identification of time and gauge fixation suggests a possible way
for solving the problem of time in quantum cosmology. This was
strongly supported by, for example, Barvinsky \cite{ba93} (see
below), and we have developed the idea for its application in the
path integral for homogeneous cosmologies (see below and
\cite{si02b}). An interesting development of this line of work
within the canonical formalism is that by Cavagli\`a, De Alfaro and
Filippov in \cite{cdf95}. In their proposal, canonical gauge fixing
is used to reduce the system: one degree of freedom is given as a
function of the remaining ones and the time parameter $\tau$, and a
true (called ``effective'' by the authors) Hamiltonian is obtained;
this Hamiltonian may in general depend on the time parameter. The
gauge choice is dictated by the simplicity of the Hamiltonian for
the reduced system. Once the reduction is performed, then the
system is quantized in the reduced canonical phase space; this is
achieved by writing a Schr\"odinger equation which is in general
$\tau$-dependent. In a given gauge, the time parameter is connected
to the canonical degree of freedom that has been eliminated.

The authors illustrate their proposal studying  a Friedmann--Robertson--Walker universe with
matter in the form of a conformal scalar field (CS) and of a SU(2)
Yang--Mills field (YM) \cite{cade94}. The corresponding Hamiltonian
constraint is
\begin{equation}
-H_{GR}+H_{CS}+H_{YM}= 0,
\end{equation}
where $H_{GR}$ is the pure gravitation Hamiltonian and
\begin{eqnarray*}
H_{CS} & = & {1\over 2}\left( p_\chi^2+V(\chi)\right)\\
H_{YM} & = & {1\over 3}\left({1\over 2}p_\xi^2+V(\xi)\right).
\end{eqnarray*}

Different gauge choices and the resulting Schr\"odinger equations
are explored. For a gauge condition in terms of the gravitational
degree of freedom like \cite{fi89} $p_\Omega +{1\over 12}e^\Omega
\cot\tau =0
$
the equation
\begin{equation}
i{\partial\over \partial
\tau}\Psi(\xi,\chi,\tau)=(H_{CS}+H_{YM})\Psi(\xi,\chi,\tau)
\end{equation}
is obtained; its solution gives a wave function for both matter
fields. A rather different choice connects the conformal field with
the time parameter: $p_\chi -\chi \cot\tau=0.
$
This gauge leads to a Schr\"odinger equation for the metric and the
Yang--Mills field:
\begin{equation}
i{\partial\over \partial \tau}\Psi(\xi,\Omega,\tau)=(H_{YM}-
H_{GR})\Psi(\xi,\Omega,\tau).
\end{equation}
An explicit solution is given for the simple case of a closed
universe with a scalar field $\phi$ with $V(\phi)=0$. The gauge
condition
$
p_\Omega-12e^\Omega\sinh\left({\tau\over\sqrt{3}}\right)=0
$
yields the equation
\begin{equation}
\left({\partial\over\partial
\tau}\mp{\partial\over\partial\phi}\right)\Psi(\phi,\tau)=0
\end{equation}
for the only physical degree of freedom $\phi$. The solutions are
of the form
\begin{equation}
\Psi(\phi,\tau)=f(\phi\pm\tau).
\end{equation}
A particular solution is
$\Psi(\phi,\tau)=Ae^{-(\phi\pm\tau)^2/2\sigma}$, which represents a
universe whose maximum probability follows the classical path
$\phi=\pm \tau$.

Apart from the usual problem of possibly non equivalent
quantizations related with different gauge choices, this procedure
has the advantage that instead of a Wheeler--DeWitt equation (even
one with a time among the coordinates), a Schr\"odinger equation is
obtained. Hence the wave function has the same properties of that
in ordinary quantum mechanics: an evolutionary form, a conserved
current and positive density. Note that the price for this
achievement has been the choice of gauges in terms of not only the
coordinates but also the momenta, so that the resulting time is in
general extrinsic.

\subsection{Avoiding inequivalent formulations}

As we have seen, the central obstruction for the existence of a
trivial correspondence between the Wheeler--DeWitt and
Schr\"odinger solutions for minisuperspaces is a constraint with a
time-dependent potential. For a class of models including some of
those studied in the preceding sections a coordinate choice
avoiding the decision between inequivalent quantum theories can be
introduced \cite{si03} . Consider the constraint (\ref{613}) and
define
\begin{eqnarray} u& =&\alpha
\exp\left({a\tilde q^1+b\tilde q^2\over 2}\right)\cosh \left(b\tilde q^1+a\tilde
q^2\over 2\right)\nonumber\\ v&
=&\alpha \exp\left({a\tilde q^1+b\tilde q^2\over 2}\right)\sinh \left(b\tilde
q^1+a\tilde q^2\over 2\right),
\end{eqnarray}
with $\alpha=\sqrt{|A|}$. These coordinates allow to write the
constraint in the equivalent (scaled) form
\be H=
-p_u^2+p_v^2+\eta m^2=0, \ee
with $\eta=sgn (A)$ and $m^2=4/|a^2-b^2|$. It is clear that
commutators cannot appear now; hence the Wheeler--DeWitt equation
is equivalent to two Schr\"odinger equations. The time is $u$ or
$v$, depending on $\eta$. The double sign given by $\eta$
corresponds to both possible sheets of the constraint surface where
the evolution can take place.

Let us illustrate this coordinate choice with some simple dilatonic
cosmologies (see \cite{si02b} and references therein); consider the
scaled constraint
$$H=-p_\Omega^2+p_\phi^2+2ce^{6\om+\phi}=0$$
which corresponds to a flat model with dilaton field $\phi$. For
$c<0$ we have $t=\pm v$, while for $c>0$ we obtain $t=\pm u$. In
the case $c<0$ (for which the dilaton $\phi$ is itself a globally
good time as $p_\phi\neq 0$), we obtain $-\infty<t<\infty$ on both
sheets of the constraint determined by the sign of $p_v$; in the
case $c>0$ (which admits $\om$ as a global time), instead, we have
that $t$ goes from $-\infty $ to $0$ on the sheet $p_u>0$, and from
$0$ to $\infty$ on the sheet $p_u<0$, with $t\to 0$ corresponding
to the singularity $\om\to -\infty$. If we include in the model a
non vanishing antisymmetric field $B_{\mu\nu}$ coming from the
$NS$-$NS$ sector of effective string theory, the constraint turns
to
$$H=
-p_\Omega^2+p_\phi^2+2ce^{6\om+\phi}+\lambda^2e^{-2\phi}= 0
$$
which in principle does not admit the proposed coordinate change.
Moreover, in the case $c<0$ the model does not admit an intrinsic
time. However, because these models come from the low energy string
theory, which makes sense in the limit $\phi\to -\infty$, then the
$e^\phi\equiv V(\phi)$ factor in the first term of the potential
verifies $V(\phi)=V'(\phi)\ll 1$, and we can replace $ce^\phi$ by
the constant $\overline c\ll c$. We can then perform the canonical
transformation introduced for the Taub universe to obtain a
constraint with only one term in the potential:
$$H=-p_\om^2+p_s^2+2\overline c e^{6\om}=0,$$
and we can apply our procedure starting from this constraint. As
before, for $\overline c<0$ we obtain $t=\pm v$, while for
$\overline c>0$ we obtain $t=\pm u$. Now, because both $u$ and $v$
depend on the coordinate $s$ which involves in its definition the
original momenta, the time is extrinsic (note that in the case
$\overline c<0$ an intrinsic time does not exist). However, in the
case $\overline c>0$, $t$ behaves with $\om$ as it did in the
absence of the antisymmetric field; $t$ goes from $-\infty$ to $0$
on the sheet $p_u>0$ of the constraint surface and from $0$ to
$\infty$ on the other sheet, while $t\to 0$ for the singularity
$\om\to -\infty$.

\subsection{Two-component wave function}

A two-component formulation is a possible way to associate a system
of differential equations which are first order in the derivative
respect to time to a Klein--Gordon type equation, as it is the
Wheeler--DeWitt one. Hence a Schr\"odinger equation is obtained, to
which well known resolution procedures can be applied and an
interpretation in terms of a well defined inner product can be
given. As it was recently shown in Refs.
\cite{mos98,mos0205,mos0209}, such idea can be effectively carried
out for some minisuperspace models. The procedure reduces the
resolution of the Wheeler--DeWitt equation to an eigenvalue problem
analogous to that of a non-relativistic harmonic oscillator and a
series of algebraic equations which can be solved by iteration. The
application of the theory of pseudo-Hermitian operators
\cite{mos0209} allows to solve the problem of constructing an
invariant positive-definite inner product on the space of solutions
of the Wheeler--DeWitt equation.

The method has been exemplified with a Friedmann--Robertson--Walker
universe with matter in the form of a massive scalar field $\phi$.
The corresponding second order equation is reduced by identifying
the logarithm of the scale factor, $\Omega$, as time variable, and
defining a wave function
\be
\Psi={1\over {\sqrt 2}}\left( \begin{array}{c}\psi+i\dot\psi \\ \psi-i\dot
\psi \end{array}
\right)
\ee
where dots mean derivatives respect to $\Omega$, and a
time-dependent Hamiltonian operator
\be
H={1\over 2}\left( \begin{array}{cc} 1+D & -1+D \\ 1-D & -1-D\end{array}
\right),
\ee
where $D=
-{\partial^2\over\partial\phi^2}-ke^{4\Omega}+m^2\phi^2e^{6\Omega}$.
This leads to a Schr\"odinger equation
\be
i\dot\Psi=H\Psi
\ee
which is solved by finding the solutions for the eigenvalue problem
$H\Psi_n=E_n\Psi_n$ (see Ref. \cite{mos0205} for the details). For
the closed $(k=1)$ model imaginary eigenvalues are obtained for
$e^\Omega>m$; the corresponding eigenvectors are null. The not
completely satisfactory feature of imaginary eigenvalues is
associated to the fact that the scale factor is really not a global
time for the closed Friedmann--Robertson--Walker universe.

\section{Beyond the minisuperspace approximation}

In this section we shall review two approaches dealing with the
full theory: the first consists in a proposal by Brown and Kucha\v
r \cite{ku95} for using dust as a time within the canonical
formalism; this leads to a constraint linear in the momentum
conjugated to the corresponding field, and therefore to a
Schr\"odinger equation; the second is the program by Barvinsky
\cite{ba86a,ba86b,ba87,ba93}, presented both in the canonical and
in the path integral formalisms; this contains the basic ideas
underlying any deparametrization procedure.

\subsection{Dust as time}

The proposal presented by Brown and Kucha\v r in Ref. \cite{ku95}
is to find a medium which, when quantizing the system in the Dirac
formulation for constrained systems, leads to a Schr\"odinger
equation --a functional one, because the proposal is presented at
the general superspace level.

It is found that incoherent dust, that is, dust which interacts
only gravitationally, is a good choice for a time variable. A
central feature of dust is that the Hamiltonian in the resulting
Schr\"odinger equation does not depend on the dust variables. Hence
the Hamiltonian density commutes (then allowing for simultaneously
defining it by spectral analysis), and the equation can be solved
by separating dust (time) from the gravitational degrees of
freedom.

The Hamiltonian and momentum constraints resulting when adding the
dust contribution in the action can be put in a form such that they
can be solved in the dust momentum field. This leads to new
constraints $H_\uparrow (X)$ and $H_{\uparrow k}(X)$, the first
generating dynamics along dust flow lines, and the others inducing
motion on the surfaces of constant proper time of dust. The true
Hamiltonian associated to the choice of dust as time is a square
root $G(X)$ depending only on the gravitational degrees of freedom.

The variables $T, Z^k$ and their conjugate quantities $M, W_k$
($k=1,2,3$) are introduced such that the values of $Z^k$ are the
comoving coordinates of the dust particles, while $T$ is the proper
time along the particle flow lines. In terms of these new variables
the Hamiltonian and momentum constraints of the whole system read
\begin{eqnarray}
H_{\uparrow}&=&P(X)+h(X,g_{ab},p^{ab})=0\\
H_{\uparrow k}& = & P_k(X)+h_k(X,T,Z^k,g_{ab},p^{ab})=0
\end{eqnarray}
where
\begin{eqnarray*}
h(X)&=&-\sqrt{G(X)},\\
G(X)&=&(H^G)^2-g^{ab}H^G_aH^G_b,\\
h_k(X)&=&Z^a_kH^G_a+\sqrt{G(X)}T_{,a}Z^a_k;
\end{eqnarray*}
here $H^G$ and $H^G_a$ are the usual Hamiltonian and momentum
constraints of pure gravitation, and $P$ is the projection of the
rest mass current of dust into the four velocity of Eulerian
observers, while $P_k=-PW_k$. Note that the Hamiltonian $h(X)$ does
not depend on the dust variables. To proceed with the canonical
quantization a new set of variables ${\bf T}(z), {\bf P}(z), {\bf
g}_{kl}(z),{\bf p}^{kl}(z)$ is introduced with the following
meaning: ${\bf T}$ is the proper time along the dust worldline
whose Lagrangian coordinate is $z^k$, ${\bf P}$ is the dust rest
mass per unit coordinate cell, and ${\bf g}_{kl}$ is the metric
giving the proper distance between nearby particles with
coordinates $z^k$ and $z^k+dz^k$. This yields the Hamiltonian
constraint
\be
{\bf H}_{\uparrow}={\bf P}(z)+{\bf h}(z,{\bf g}_{kl},{\bf p}^{kl})=0.
\ee
Hence the resulting Schr\"odinger equation for the wave functional
${\bf \Psi}(Z,{\bf T},{\bf g})$ is
\be
i{\delta {\bf \Psi}(Z,{\bf T},{\bf g})\over \delta {\bf T}}= {\bf h}(z,{\bf
g},{\bf p}){\bf \Psi}(Z,{\bf T},{\bf g}).
\ee
But because the wave functional must fulfill the momentum
constraints which as operators are functional derivatives respect
to the canonical coordinates $Z^k$, then ${\bf \Psi}$ does not
depend on $Z^k$ and hence it is obtained
\be
i{\delta {\bf \Psi}({\bf T},{\bf g})\over \delta {\bf T}}= {\bf h}(z,{\bf
g},{\bf p}){\bf \Psi}({\bf T},{\bf g}).
\ee
The replacement of the Wheeler--DeWitt equation by this functional
Schr\"odinger equation thus allows to define a conserved positive
definite inner product. (To be precise, to obtain a self adjoint
new physical Hamiltonian ${\bf h}$, the theory must be restricted
to the subspace given by the positive eigenvalues of the operator
$\hat {\bf G}$ associated to $G$ defined above). It is interesting
to note that the idea by Brown and Kucha\v r has been recently
applied by A. Sen \cite{sen} to string cosmology, with the tachyon
playing the same role as dust, that is, as time variable for the
system and leading to a Schr\"odinger equation.

\subsection{The reduction to true degrees of freedom as a way to  unitary
quantum cosmology}

A central contribution in the search for a unitary quantum theory
for gravitation has been that by Barvinsky
\cite{ab96,bapo86,ba86a,ba86b,ba87,ba93}. It is clearly beyond the scope
of these notes to give a thorough review of his seminal works
--moreover, they are mostly devoted to the full theory, and not to
the minisuperspace approximation--, but because most of our own
contributions have largely drown on them, here we shall briefly
comment some of their central aspects.

The program starts from the reduction of gravity theory to true
physical variables $\zeta$ by means of gauge fixation, which
appears natural after the following considerations: The dynamical
evolution, which includes the problem of the multiplicity of times
associated to the fact that the separation between successive
three-hypersurfaces is arbitrary, can be reproduced by gauge
transformations \cite{ba93}. The extremal condition $\delta S=0$
gives the canonical equations
\begin{equation}
{\frac{dq^{i}}{d\tau }}=N_\mu[q^{i},{\cal H^\mu}],\ \ \ \ \ \ \ \ \
{\frac{dp_{i}}{
d\tau }}=N_\mu[p_{i},{\cal H^\mu}].\label{evol}
\end{equation}
The solution of these equations describes the evolution of a
spacelike hypersurface along the timelike direction, and the
presence of the multiplier $N$ introduces an arbitrariness in the
evolution which is associated to a multiplicity of times. From a
different point of view, the constraint ${\cal H}= 0$ acts as
a generator of gauge transformations which can be written
\begin{eqnarray}
\delta _{\epsilon }q^{i}& = & \epsilon_\mu (\tau )[q^{i},{\cal H^\mu}],\nonumber\\
\delta
_{\epsilon }p_{i}& = & \epsilon_\mu (\tau )[p_{i},{\cal H^\mu}],\nonumber\\ \delta
_{\epsilon }N_\mu & = & {\frac{\partial \epsilon_\mu (\tau )}{\partial \tau }}-
u_\mu^{\nu\rho}\epsilon_\rho N_\nu,\label{gauge}
\end{eqnarray}
where $u_\mu^{\nu\rho}$ are the structure functions of the constraints algebra.
Then, from (\ref{evol}) and (\ref{gauge}) we see that the dynamical
evolution can be reproduced by a gauge transformation progressing
with time, that is, any two successive points on each classical
trajectory are connected by a gauge transformation; this leads to
the idea of identifying time and true degrees of freedom by fixing
the gauge.

Once the gauge is fixed, the constraints are solved, then yielding
a true non vanishing Hamiltonian and a physical time. The reduced
system is then quantized, and the theory is reformulated in terms
of the initial superspace variables $q$ (that is, in terms of the
canonical variables including spurious degrees of freedom). This
procedure allows to obtain a wave function $\Psi(q)$ solving the
operator form of the constraints and including the central feature
of a precise inner product allowing for a clear probabilistic
interpretation.

After the reduction of the classical theory the quantization
follows the usual path integral procedure; and the subsequent
reformulation in terms of the original variables gives a unitary
gauge-independent superspace propagator which allows to evolve
initial conditions on a given Cauchy surface, that is, to evolve
from a given subspace in superspace. This allows for the obtention
of a wave function $\Psi(q)$ for which the measure is well defined
in the sense that the probability amplitude is conserved in the
superspace theory including a multiplicity of times.

Within this context, Barvinsky has also analysed the generalized
Batalin--Fradkin--Vilkovisky (BFV) canonical quantization and has
shown that the superspace wave function $\Psi(q)$ is an
intermediate step between the wave function for the physical
degrees of freedom $\psi(\zeta)$ and the quantum states in the
extended BFV Hilbert space. Also, a definite operator ordering, and
the corresponding quantum corrections, are given which ensure the
closure of the constraint algebra and their hermiticity properties
resulting from the BFV formalism.

Some points of Barvinsky's proposal can be outlined: 1) The absence
of an asymptotically free limit in gravity theory forces the choice
of a coordinate representation; this implies the restriction to
systems which admit an intrinsic time, which appears in the
reduction procedure as a result of imposing gauges not involving
the momenta. 2) To avoid a frozen formalism, the appropriate gauge
conditions must be explicitly time-dependent:
\be \chi(q,\tau)=0.\ee
3) Because of the form of the Hamiltonian constraint, which is
quadratic in the momenta, the theory in the reduced space described
by the set of  canonical variables $(\zeta^A,\pi_a)$ includes {\it two}
physical Hamiltonians $H_\pm$ fulfilling
\be H_-(\zeta,-\pi,\tau)=-H_+(\zeta,\pi,\tau),\ee
corresponding to two disjoint theories. 4) It is assumed that the
quantum description in terms of the physical degrees of freedom
$(\zeta^A,\pi_a)$ is the gauge invariant quantum theory for the
original variables $(q^i,p_i,N^\mu)$. 5) The theory in the physical
subspace is given by the commutation relations
\be
[\zeta^A,\pi_A]=i\delta^A_B,\ee
and the Schr\"odinger equation
\be
i{\partial\over\partial \tau}\psi
(\zeta,\tau)=H(\zeta,\pi,\tau)\psi(\zeta,\tau)\ee
with the inner product
\be
\langle\tilde\psi\vert\psi\rangle=\int d\zeta\,\tilde\psi^*\psi,\ee
or, in the path integral formulation, by the propagator
\be
K(\zeta_2,\tau_2\vert\zeta_1,\tau_1)=\int D\zeta
D\pi\exp\left(iS[\zeta,\pi]\right).\ee
6) Once the reduced theory
has been constructed, so ensuring the unitarity of the quantum
description, a reformulation in terms of the original variables is
performed. This means the obtention of a wave function $\Psi(q)$
and a gauge fixing in superspace establishing the correspondence
between $\Psi(q)$ and $\psi(\zeta)$, and the definition of a
conserved inner product in the Hilbert space associated with
superspace, as well as the proof of the consistency of different
gauge choices. The existence of two disjoint theories at the level
of the true degrees of freedom is reflected in the fact that {\it
two} superspace propagators $${\bf K}^+(q_2\vert q_1)\ \ \ \ \ \ \
\ {\bf K}^-(q_2\vert q_1)$$
are obtained. Because these propagators are gauge-independent,
after the transition from the theory for the true degrees of
freedom to the unitary theory in superspace one obtains a wave
function $\Psi(q)$ which depends only on the initial gauge
conditions, which are included in the initial-value data for it
(see Ref. \cite{ba93} for the details).

\end{document}